\documentclass[reprint,twocolumn,notitlepage,prb,superscriptaddress,showpacs]{revtex4-2}

\usepackage[usenames,dvipsnames]{color}

\usepackage{comment}
\usepackage{graphicx}
\usepackage[utf8x]{inputenc}
\usepackage{lineno}
\usepackage{color}
\usepackage{hyperref}
\usepackage{amsmath}
\usepackage{braket}
\usepackage{caption}
\usepackage{subcaption}
\usepackage{amsfonts}

\begin{document}

\title{Renormalized Perturbation Theory for Fast Evaluation of Feynman Diagrams on the Real Frequency Axis}
\author{M. D. Burke}
\affiliation{Department of Physics and Physical Oceanography, Memorial University of Newfoundland, St. John's, Newfoundland \& Labrador, Canada A1B 3X7} 
\author{Maxence  Grandadam}
\affiliation{Department of Physics and Physical Oceanography, Memorial University of Newfoundland, St. John's, Newfoundland \& Labrador, Canada A1B 3X7}
\author{J. P. F. LeBlanc}
\email{jleblanc@mun.ca}
\affiliation{Department of Physics and Physical Oceanography, Memorial University of Newfoundland, St. John's, Newfoundland \& Labrador, Canada A1B 3X7}

\date{\today}
\begin{abstract}
We present a method to accelerate the numerical evaluation of spatial integrals of Feynman diagrams when expressed on the real frequency axis.  This can be realized through use of a renormalized perturbation expansion with a constant but complex renormalization shift.  The complex shift acts as a regularization parameter for the numerical integration of otherwise sharp functions.  This results in an exponential speed up of stochastic numerical integration at the expense of evaluating additional counter-term diagrams.  We provide proof of concept calculations within a difficult limit of the half-filled 2D Hubbard model on a square lattice.
\end{abstract}

\maketitle
\section{Introduction}
The Matsubara formalism is the dominant representation of finite-temperature many-body physics and is foundational for solving correlated electron problems at finite temperatures.\cite{benchmarks,Schaefer:2020,jia:2020,hubb:review:theory,hubb:review:comp} 
This formulation is used in both perturbative diagrammatic methods as well as non-perturbative methods that are typically formulated in imaginary times rather than on the Matsubara frequency axis.\cite{vanhoucke:fixed,kozik:2010,georges:1996,gull:revmod}  
Central to the textbook utility of the Matsubara formalism is the ability to 
obtain results on the real-frequency axis.  This process, known as analytic continuation, is equivalent to simply replacing the Matsubara frequency $i\omega_n\to \omega+i\Gamma$, which is exact in the $\Gamma\to 0^+$ limit.  
Despite the conceptual simplicity, numerical methods are typically not able to implement analytic continuation.  Instead, computational methods that obtain observables for a discrete set of  $i\omega_n$ (or imaginary times $\tau$) must invoke an auxilliary 
procedure, so-called numerical analytic continuation.  
A number of schemes for numerical analytic continuation exist such as: 
maximum entropy inversion MAXENT\cite{jarrell:maxent,maxent},
Pade approximants\cite{pade}, or more modern methods such as the Carath\'eodory and Nevanlinna algorithms\cite{Cara:maxent,nevanlinna:maxent} 
as well as machine learning approaches.\cite{ML:fournier,ML:millis,ML:yoon}  

All of the methods of numerical analytic continuation are based upon the inversion of an ill-posed problem and produce one of a potentially infinite number of solutions.  While there has been significant progress in constraining these solutions, the initial ill-posed nature of the problem cannot be escaped.  In the case of diagrammatic monte carlo (DiagMC) the need for numerical analytic continuation has been alleviated through automated analytic approaches to the Matsubara summation.  A handful of such algorithmic approaches exist as well as full analytic solutions to some diagrammatic expansions in imaginary time.\cite{ami,git,libami,jaksa:2020,jaksa:analytic}  
Together these methods can be referred to as real-frequency diagrammatic Monte-Carlo (RF-DiagMC).
The advantage of RF-DiagMC is that the key component, the Matsubara summation, is treated analytically and this allows for direct symbolic replacement $i\omega_n\to \omega+i\Gamma$.  

Perhaps the most successful incarnation of RF-DiagMC utilizes a method called algorithmic Matsubara integration (AMI) that allows one to automatically (and virtually instantaneously) obtain solutions to the Matsubara sums of a wide class of Feynman diagrams.\cite{ami,libami}  It has been applied to diagrammatic expansions of single particle properties (Green's functions, self-energies, and densities),\cite{mcniven:2021} charge and spin susceptibilities,\cite{mcniven:2022} screened interactions,\cite{igor:spectral} and others and is valid for any frequency independent interaction that can be formulated in momentum space such as the Hubbard model and also both the direct Coulomb and the screened Yukawa potentials.\cite{igor:james,igor:spectral} 

The result of the AMI procedure is an analytic expression that must still be summed over remaining internal degrees of freedom - typically spatial variables such as internal momentum or band-indices in the case of multi-band problems. This is no different from classic DiagMC methods with the exception that, using AMI, external frequency variables can be evaluated \emph{either} on the Matsubara axis or the real-frequency axis.  Although AMI provides an apparent solution to the analytic continuation problem, there remains a largely unexplored but fundamental issue, namely that obtaining rigorously correct results via analytic continuation hinges on evaluation of an integrand in the $\Gamma\to0^+$ limit.   While in principle $\Gamma$ can be made arbitrarily small, in practice $\Gamma$ plays a role of a complex numerical regulator in the integration of remaining spatial variables and must always remain finite.  The impact of a non-zero $\Gamma$ on observables is a broadening of sharp features occurring on a frequency range $\Delta\omega<\Gamma$ as well as broadening that is similar to thermal effects, necessitating that $\Gamma << T$.  

For low-order diagrams the dimensionality of spatial integrals is low, and the variance of the integrand typically scales $\propto \Gamma^{-1}$ in which case the role of $\Gamma$ can be controlled and reliable results can be obtained.  Such is typically the case in real frequency evaluation of the LDA+GW method for material calculations.  There one primarily evaluates low dimensional integrals for Lindhard function in the RPA expansion  
of $W$ followed by a low dimensional integration of the self-energy approximated by $\Sigma = GW$.  
However, to extend GW methods to include higher order corrections to the self-energy then we expect that the integrand for order $m$ diagrams will contain peaks that scale as $\Gamma^{-m}$ which by nominal order in $m$ ($m=5$ or $6$) can cause numerical overflow for standard or double floating point precision arithmetic.  What is worse is that when performing integration via Monte-Carlo methods, where the uncertainty is guaranteed to scale as $\sigma/\sqrt{N}$ for a number of samples $N$, the variance $\sigma$ becomes larger with decreasing $\Gamma$ and is infinite in the $\Gamma \to 0^+$ limit even if there is no numerical sign-problem (the integrand is sign-definite).  Worse still, as the dimensionality increases the variance for fixed value of $\Gamma$ increases dramatically.  Together this leads to a scaling of the variance $\sigma \propto \Gamma^{-\gamma}$ where $\gamma$ represents an effective dimensionality\cite{eff_dim} that is integrand dependent.  This reintroduces the dimensionality constraint in uncertainty achieved via Monte-Carlo integration making it virtually impossible to extend RF-DiagMC methods to high enough order to draw concrete conclusions on perturbative problems in the true $\Gamma\to0^+$ limit. 

In this paper we address this issue through a renormalized perturbative approach.  Renormalized perturbative methods have widely been used for strongly correlated electron systems, such as the 2D and 3D Hubbard models,\cite{lenihan:2022,simkovic2017determinant} within non-symbolic diagrammatic Monte-Carlo methods, such as connected Determinant Monte Carlo (cDET)\cite{rossi2017determinant}, with the intent of expanding the radius of convergence of the perturbative expansion. Unlike those works, we primarily study the impact of a constant complex shift, $z=i\alpha$, whose imaginary part, $\alpha$, masquerades as a numerical regulator for performing the spatial integrations but for which, unlike $\Gamma$, its impact on numerical results can be systematically removed.  This comes at the expense of evaluating an infinite set of Feynman counter-term diagrams.  We show that, despite the additional counter-terms, this approach exhibits a massive computational advantage and may provide access to the $\Gamma\to 0^+$ limit that cannot be accessed otherwise.

\section{Model and Methods}
\subsection{Hubbard Hamiltonian and Parameters}
We study the single-band Hubbard Hamiltonian on a 2D square lattice\cite{benchmarks},
\begin{eqnarray}\label{E:Hubbard}
H = \underbrace{\sum_{ ij \sigma} t_{ij}c_{i\sigma}^\dagger c_{j\sigma}}_{H_0} + \underbrace{U\sum_{i} n_{i\uparrow} n_{i\downarrow}}_{H_v},
\end{eqnarray}
where $t_{ij}$ is the hopping amplitude, $c_{i\sigma}^{(\dagger)}$ ($c_{i\sigma}$) is the creation (annihilation) operator at site $i$, $\sigma \in \{\uparrow,\downarrow\}$ is the spin, $U$ is the onsite Hubbard interaction, $n_{i\sigma} = c_{i\sigma}^{\dagger}c_{i\sigma}$ is the number operator.  We restrict the sum over sites to nearest neighbors, resulting in the free particle energy dispersion 
\begin{eqnarray}
\nonumber\epsilon(\textbf k)=-2t[\cos(k_x)+\cos(k_y)]-\mu,
\end{eqnarray} 
where $\mu$ is the chemical potential, and $t$ is the nearest neighbor hopping amplitude.  Throughout, we work with energies in units of the hopping, $t=1$. We restrict calculations to the half-filled problem $\mu=0$ since it is the most computationally challenging for direct perturbative methods.

\subsection{Renormalized Diagrams and Counter-Terms}
The renormalized perturbative expansion involves introducing a constant single-particle term into our Hamiltonian $\delta=z\sum\limits_{i\sigma}\hat{n}_{i\sigma} $.  In order to not modify the original Hamiltonian we write
\begin{align}
    H&=H_0+H_v + \delta - \delta \\
    &= (H_0-\delta) + (H_v + \delta)\\
    &= H_0^\prime + H_v^\prime .
\end{align}
Since we have made no change to the Hamiltonian, we are free to expand around the known solution to $H_0^\prime$ with respect to $H_v^\prime$.  The role of the complex shift, $z$, is equivalent to an effective shift in chemical potential such that the new non-interacting Green's function on the Matsubara axis is given by
\begin{equation}\label{eqn:G}
    G_0^{-1}(k,i\omega_n)=i\omega_n - \epsilon_k +\mu +z .
\end{equation}

In order to compensate for the inclusion of $z$ in the bare propagator, the expansion of $H_v^\prime$ will, in comparison to an expansion of $H_v$, spawn an infinite set of counter-term diagrams that represent self-energy insertions with amplitudes give by powers of the $z$ correction. In principle if all counter-terms are included to infinite order in $z$ then the result will be independent of the choice of $z$.  Considering the expansion of the self energy one would obtain the full self energy via 
\begin{equation}
    \Sigma_k(i\omega_n)=\sum\limits_{n=0}^{\infty} \sum\limits_{s=0}^{\infty} a_{n,s}(z) U^n(z)^s.
\end{equation}
While the coefficients at each order in $U$ and $z$, $a_{n,s}$ are dependent upon the choice of $z$ the infinite summation is not.  In practice however, one will obtain results only up to a particular truncation order in both $U$ and $z$ in which case
\begin{equation}
    \Sigma_k^{(m,c)}(i\omega_n,z)=\sum\limits_{n=0}^{m} \sum\limits_{s=0}^{c} a_{n,s}(z) U^n(z)^s.
\end{equation}
In this case the resulting self-energy with truncation at order $m$ in $U$ and order $c$ in $z$ is no longer fully independent of $z$.  If we take a purely imaginary $z=i\alpha$ and perform analytic continuation by replacing $i\omega_n \to \omega + i\Gamma$ then we see that each resulting Green's function, Eq.~(\ref{eqn:G}), is dependent upon two numerical regulators; a fundamental regulator, $\Gamma$, that must be non-zero but whose value must be much less than any other scale, and a second regulator, $\alpha$, whose effect can be systematically removed by increasing the cutoff in counter-term order, $c$.

In practice, one absorbs both the chemical potential, $\mu$, and this new regulator $\alpha$ into the dispersion.  While each Green's function includes only a single instance of $\epsilon_k$, the integrand after processing with AMI will contain linear combinations of $\epsilon_{k_i}$.  As a result, some terms will not benefit from the regulation procedure in the Green's functions, and hence a small non-zero value of $\Gamma$ is always required.  

\subsection{Integration Methodologies}
The number of integration dimensions is a typical metric for the level of difficulty evaluating an integral numerically.  However, if integrands are unstructured then the inclusion of additional dimensions does not negatively impact stochastic estimation processes.  This leads to the concept of effective dimensionality which is borrowed from the study of Quasi-monte carlo (QuasiMC) integration methods.\cite{eff_dim}  QuasiMC takes advantage of low discrepancy number sequences that optimally span the integration space while respecting the central limit theorem.  Those methods can be shown to scale as $\log(N)^s/N$ for an $s$-dimensional integral.  Contrasting this to normal monte-carlo methods that scale as $1/\sqrt{N}$ one expects that even nominal dimension the Quasi-MC scheme should become inferior.  However, this is typically not the case in practice where one finds that Quasi-MC methods typically outperform MC for many model integrands up to extremely high-dimensional integrals.  Explaining this has led to the reinterpretation of $s$ as an effective dimensionality.  In short, the effective dimensionality is a measure of how a function's value along one integration axis is impacted by variation of the other integration variables.  For many integrands the effective dimensionality is substantially smaller than the actual dimensionality.  
Turning to the present work, we have a scenario where the effective dimensionality is tied to the value of the regulator $\Gamma$ since it controls the sharpness of the functions being integrated.    

The most well known issue in Monte-Carlo sampling is the existence of a sign problem, where the average of a function is small due to sign changes in the integrand.  This causes the variance to be large and while the scaling of monte carlo methods remains $\propto 1/\sqrt{N}$, the prefactor, $\sigma$, becomes large making it impossible to obtain a reliable result.  
This is not the only way to generate a large variance, in particular integrands that are sparse with sharp peaks result in similarly large variance that typically worsens as the dimensionality of the integrand increases.  In those cases one must target to reduce the variance specifically.  Many schemes exist to do this such as the Metropolis-Hastings algorithm, as well as importance sampling with adaptive grids.  Regardless of the use of variance reduction methods, real frequency integrands cannot be evaluated in the $\Gamma \to 0^+$ limit.  

Throughout we will present results using only naive monte-carlo (sampling with a flat distribution) because it is both the simplest to implement but also since it has no variance reduction it remains unbiased and stands to benefit the least from our renormalization broadening $\alpha$. 

\subsection{Scaling of counter-term expansion}
Determining if such a renormalized scheme is computationally useful is dependent upon the increase in complexity weighed against the computational advantage of including a larger regulator.  For every root diagram (diagram with s=0) with a number of Green's function lines, $N_g$, when summed from $s=0\to c$ will result in a total number of diagrams equal to $N_D={N_g+c \choose N_g }$.  The computational expense is therefore increased by at least a factor of $N_D$.  
The tradeoff comes when considering the ratio of scaling of the $\alpha=0$ case leading to an overall scaling proportional to  $N_D \left[ \frac{\Gamma}{\Gamma+\alpha}\right]^{\gamma}$.  We note that in the desired $\Gamma \to 0^+$ limit this computational advantage is potentially massive so long as the effective dimension, $\gamma$, of the integrand is not zero (the variance remains independent of the dimensional scaling as would be true for a flat function).

To proceed, we generate the necessary counter-term diagrams for each diagram of interest.  We depict a handful of these at second and fourth order in Fig.~\ref{fig:diags} upon which we will base discussion.  Each counter-term insertion comes with a factor of $z$ but adds also an additional Green's function to the diagram and is therefore of slightly higher complexity.  We will focus on two root diagrams, a second order self-energy diagram and a similar fourth order diagram. 

\section{Results}
\subsection{Application at Low Order - Second Order Self Energy}
\begin{figure}
    \centering
    \includegraphics[width=0.8\linewidth]{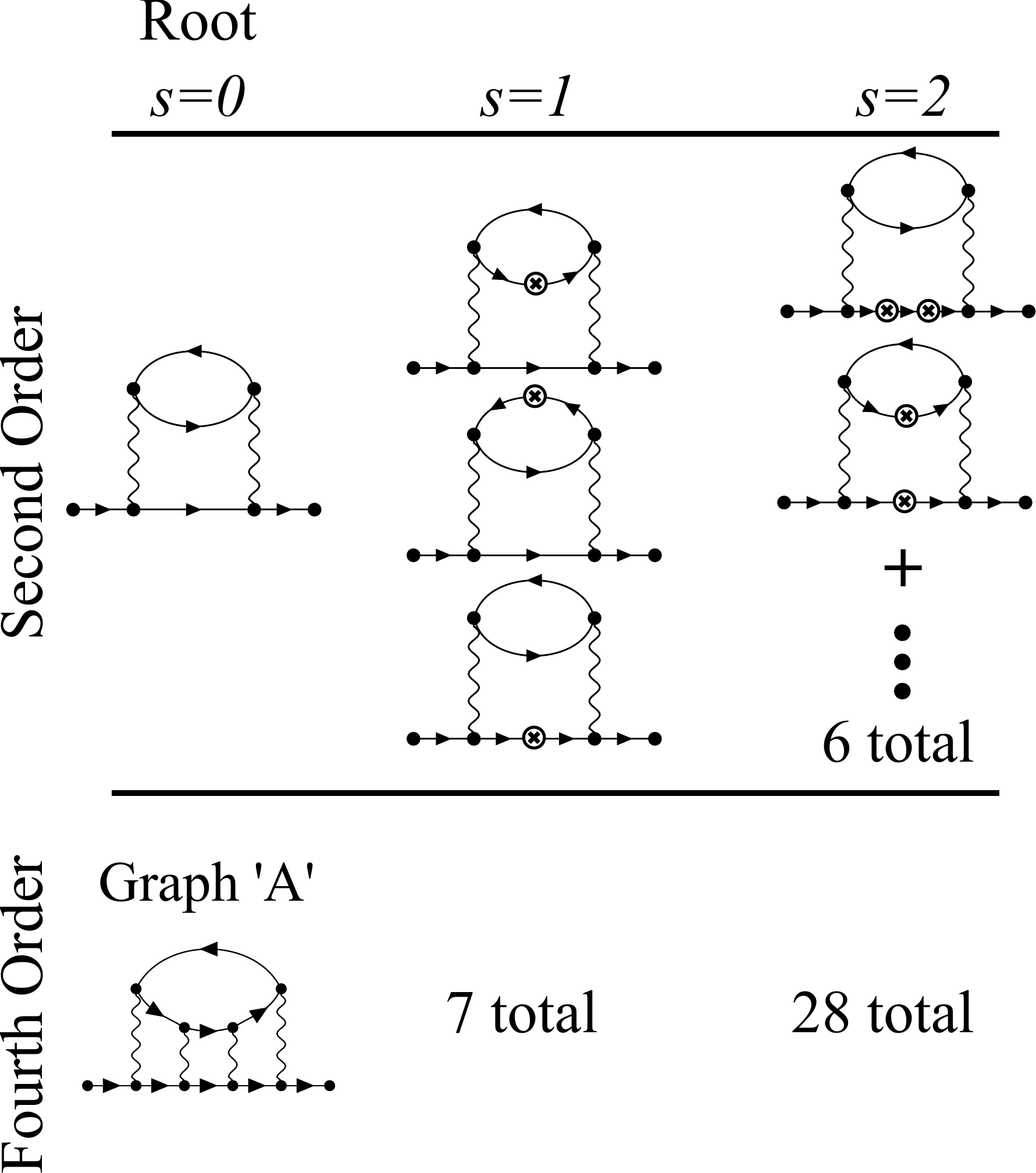}
    \caption{Root diagrams, $s=0$, at second and fourth order and example counterterm diagrams for $s=1$ and $s=2$.}
    \label{fig:diags}
\end{figure}

\begin{figure}
    \centering
    \includegraphics[width=\linewidth]{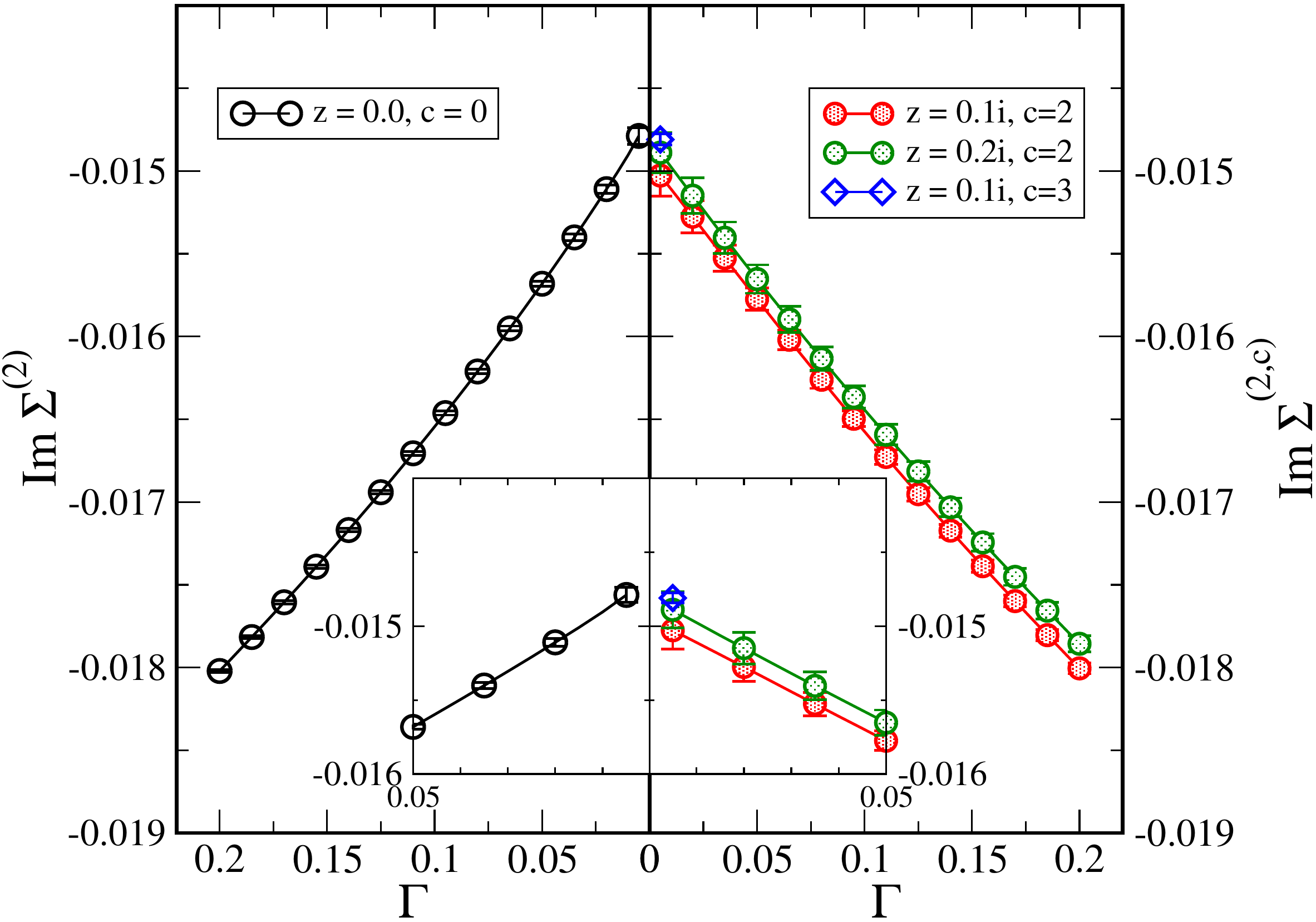}
    \caption{Imaginary part of self energy truncated to all second order diagrams up to $c$ insertions of various values of $z$ as a function of $\Gamma$. On the left-hand side, the regular approach of no insertions with $z=0$ is shown for comparison to the renormalized-perturbative approach on the right-hand side with various $c$ and $z$. Parameters of the model are: $U/t=1$, $\beta t=5.00$, $\textbf{k}=(\pi, 0)$, $\omega/t = 0.3$.}
    \label{fig:o2_gamma}
\end{figure}

As a proof of concept we examine an easily obtained result of the second order self-energy depicted in Fig.~\ref{fig:diags} evaluated for $\beta t=5$, $k=(\pi,0)$, for frequency $\omega/t=0.3+i\Gamma$ at $U/t=1$.  At low order the $\Gamma \to 0^+$ limit remains numerically tractable providing easy access to benchmark values for our truncated renormalized approach.  
Shown in the left frame of Fig.~\ref{fig:o2_gamma} is the evaluation of the root second-order diagram with $z=0$ plotted as a function of 
the analytic continuation parameter, $\Gamma$, as it is reduced at fixed computational expense.  This represents the physically correct results and we see that the error bars grow as 
$\Gamma$ is reduced.  One can easily extrapolate such data to the $\Gamma\to 0^+$ limit directly.  From this low-order contribution 
we can see clearly the importance of taking the $\Gamma \to 0^+$ limit since the result varies by $\approx 20\%$ over 
the $\Gamma=0.2 \to 0$ range. In the right-hand frame of Fig.~\ref{fig:o2_gamma} we show renormalized perturbative results for the same 
root diagram but now with finite $z$ and including also the counter-term diagrams up to cutoff order $c=2$ for $z=0.2i$ and to both $c=2$ and $3$ at 
$z=0.1i$.  Since the root diagram has three Green's functions, a total of $N_D={3+c \choose 3}$, or $N_D=10$ and $20$ for $c=2$ and $c=3$ 
respectively, are included. 
We note that these renormalized results are expected to have the same values as a function of $\Gamma$ if summed to all orders in $s$.  For this particular case, there is little computational advantage for having computed these counter-terms with the exception that one can avoid extrapolating the result to $\Gamma=0^+$ and instead choose an extremely small $\Gamma$ with relatively large $\alpha$ and systematically remove the impact of $\alpha$ by including more counter-terms. 

\begin{figure}
    \centering
    \includegraphics[width=\linewidth]{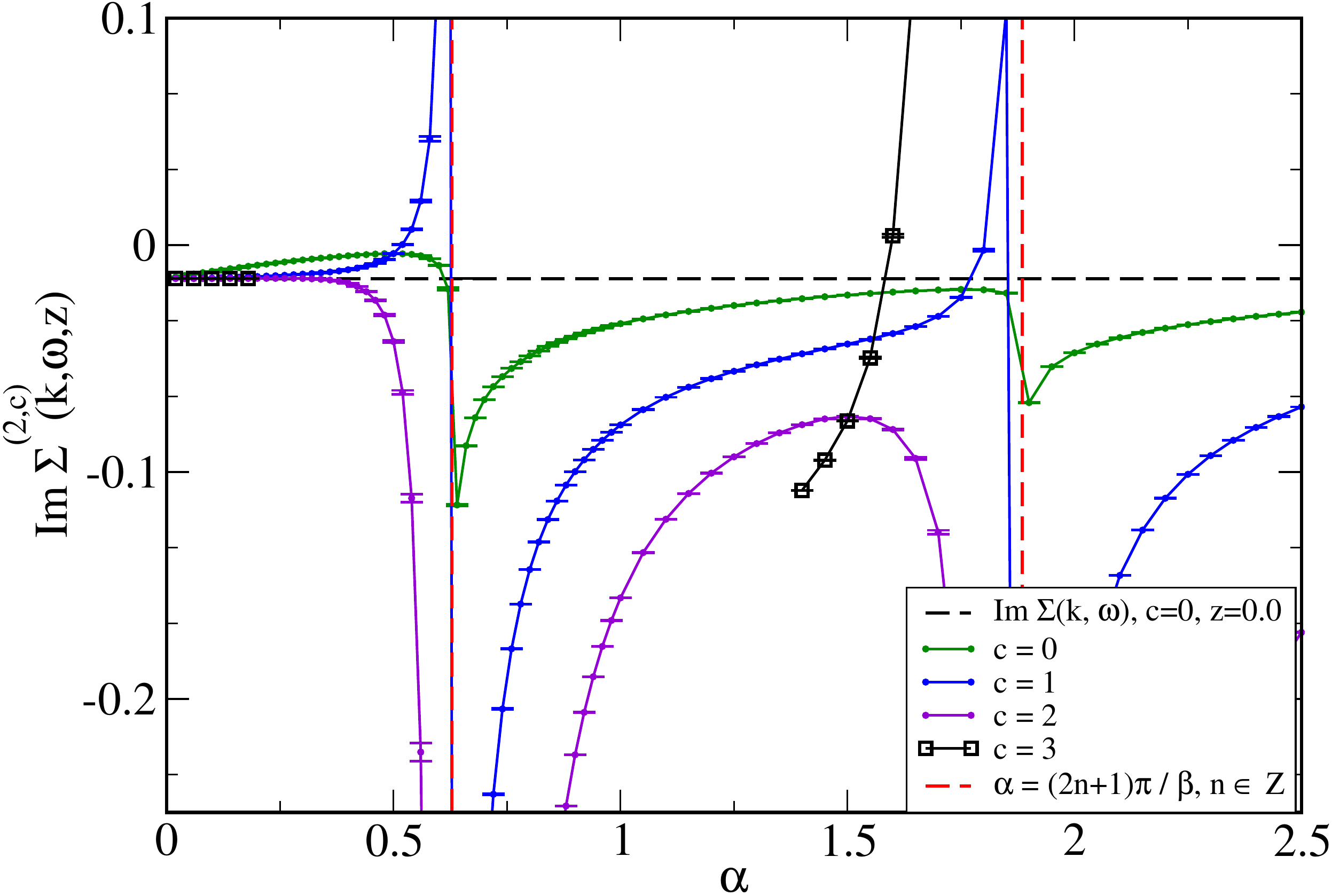}
    \caption{Imaginary part of the self energyfor the second order root diagram up to $c$ insertions as a function of $\alpha$, the magnitude of the purely imaginary renormalization shift, $z=i\alpha$. The values of the fermionic Matsubara frequencies, $\alpha=\omega_n, n=0,1$ are marked with vertical dashed-red lines. Parameters of the model are: $U/t=1$, $\beta t=5.00$, $\textbf{k}=(\pi, 0)$, $\omega/t = 0.3$, $\Gamma/t= 0.001$.  The benchmark value of ${\rm Im}\Sigma(k,\omega)$ is shown as a horizontal dashed line.}
    \label{fig:o2_alpha}
\end{figure}

In principle, one should obtain an identical result independent of the choice of $\alpha$ so long as enough counter-terms are included.  In some renormalized expansions extremely large values of $z$ are allowable.  Here however, we will see that the use of the renormalized perturbation theory along with RF-DiagMC using AMI that there are stringent requirements on the amplitude of $z=i\alpha$.  These have been mentioned previously\cite{jaksa:analytic} in that the radius of convergence in $z$ is restricted to $|z|<i\omega_0$ where $i\omega_0$ is the first Fermionic Matsubara frequency.  The reason for this limitation is due to the analytic expressions generated by AMI being incorrect when $z=i\omega_n$ unless special care is taken to account for this.  
We demonstrate the result of a fixed frequency $\omega/t=0.3$ in Fig.~\ref{fig:o2_alpha} where we plot ${\rm Im}\Sigma^{(2,c)}$ as a function of $\alpha$ for cutoff order $c=0,1,2,$ and 3.  We see that for choices of $\alpha$ near $i\omega_n$ (red vertical dashed lines) the result is clearly non-convergent.  In particular, for values of $\alpha$ between $i\omega_0$ and $i\omega_1$ the result is not-divergent but increasing the cutoff, $c$, gives a result that is systematically further from the correct benchmark value, the horizontal dashed line.
This is not the case at small $\alpha$ values, we see that all $c=1,2,$ and 3 are virtually flat for $\alpha<0.25$.  We conclude then that a range of small values of $\alpha$ provides an easy check for the impact of the truncation order $c$ allowing us to produce reliable results in the small $\Gamma$ limit, here shown for $\Gamma/t=1\times 10^{-3}$.

  We can therefore compute $\Sigma^{(n,c)}$ to an appropriate truncation order in $c$ and see that higher order counterterms do not contribute.  If one cannot access large values in $c$ (too many diagrams), one can simply reduce (or increase) the value of $\alpha$ to assess the level of accuracy.  The best choice of $\alpha$ will be a value as large as possible to take advantage of the broadening while the accessible counter-terms remain small.

\subsection{Higher order Application - Fourth Order Self Energy}
\begin{figure}
    \centering
    \includegraphics[width=\linewidth]{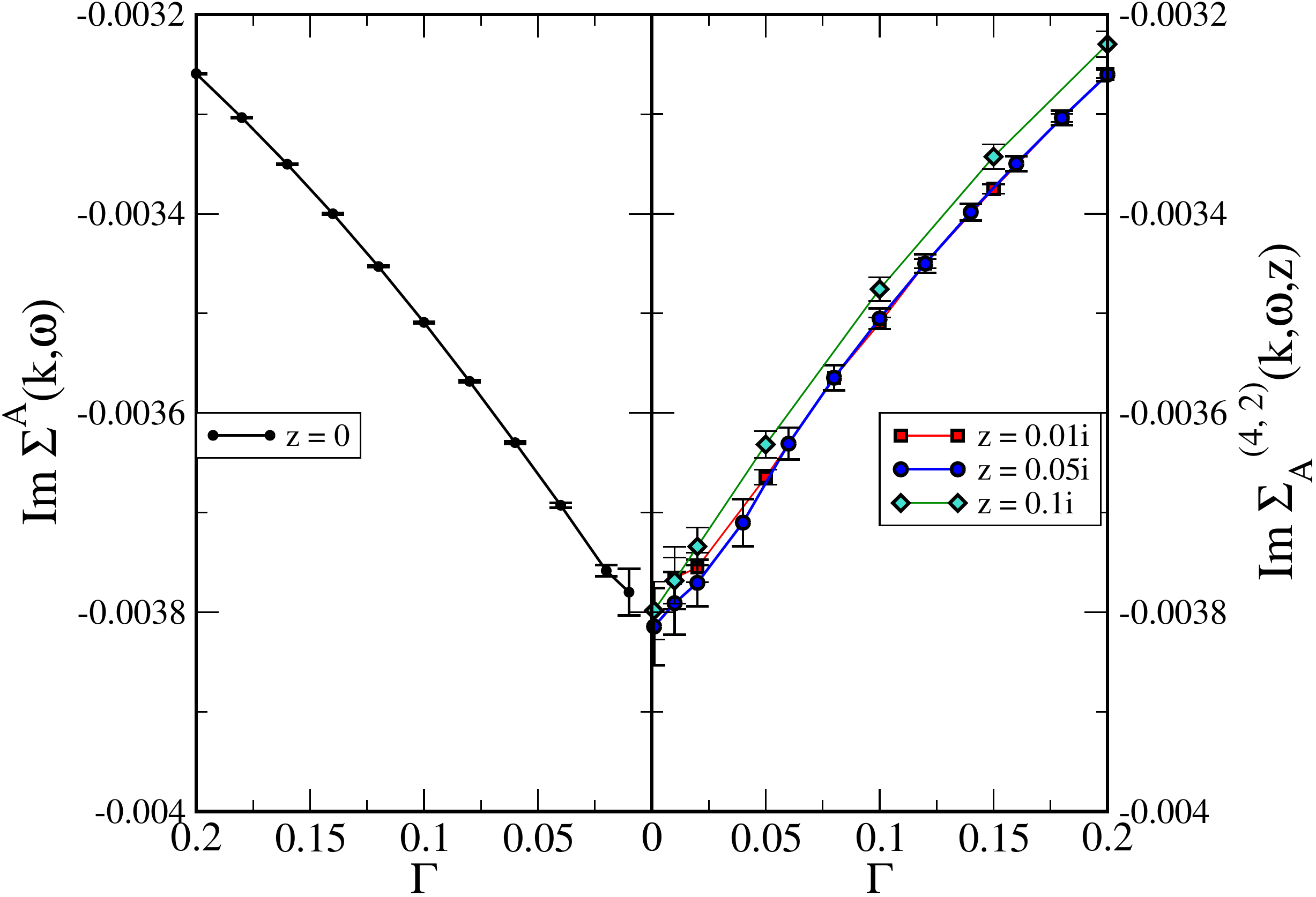}
    \caption{Evaluation of the imaginary part of the self energy on graph A as a function of $\Gamma$. Right-hand side: Summation up to $2$ insertions for various $z$. Left-hand side: the regular approach with no counterterms and $z=0$. Also shown is the fitted value of $\Gamma=0$ from the . Parameters of the model are: $U/t=1$. $\beta t=5.00$, $\textbf{k}=(\pi, 0)$, $\omega/t = 0.1$.}
    \label{fig:o4_gamma}
\end{figure}

We now focus on a much more difficult example of one fourth order root diagram, graph A (see Fig.~(\ref{fig:diags})), that spawns seven counterterms at $s=1$ and 28 counterterms at $s=2$.
We first examine the variation of this diagram including up to two counterterms.  Results are shown in Fig.~\ref{fig:o4_gamma} for $\omega=0.1+i\Gamma$  over the range $\Gamma=0.2\to0^+$ with different choices of purely imaginary $z$.
The left frame of Fig.~\ref{fig:o4_gamma} shows only the root diagram as a benchmark while the right hand frame shows the result for ${\rm Im}\Sigma_A^{(4,2)}(k,\omega+i\Gamma,z)$.  We see that in the range of $\Gamma$ shown, for this particular frequency of $\omega/t=0.1$ the variation in result is on the scale of $\approx 18\%$. On the right-hand frame, $\textrm{Im}\Sigma_A^{(4,2)}$ is calculated by summing graph A up to 2 counter-term insertions for $z=0.01i, 0.05i,$ and $0.1i$.  This demonstrates the correctness of the method since all $z$ values approach the same $\Gamma \to 0^{+} $ limit. Here we see that due to the number of diagrams computed the computational advantage, if it exists, is only for very small values of $\Gamma$.  We will probe this issue further in Section~\ref{sec:comptime}.

\subsection{Real frequency evaluation}
We explore our proposed renormalized method by calculating the contribution to the self energy from graph A as a function of real frequency $\omega$.  To illustrate the computational impacts of both $\Gamma$ and $z$ we maintain a fixed evaluation time for figure frames from left to right. 
\begin{figure*}
    \centering
    \includegraphics[width=\linewidth]{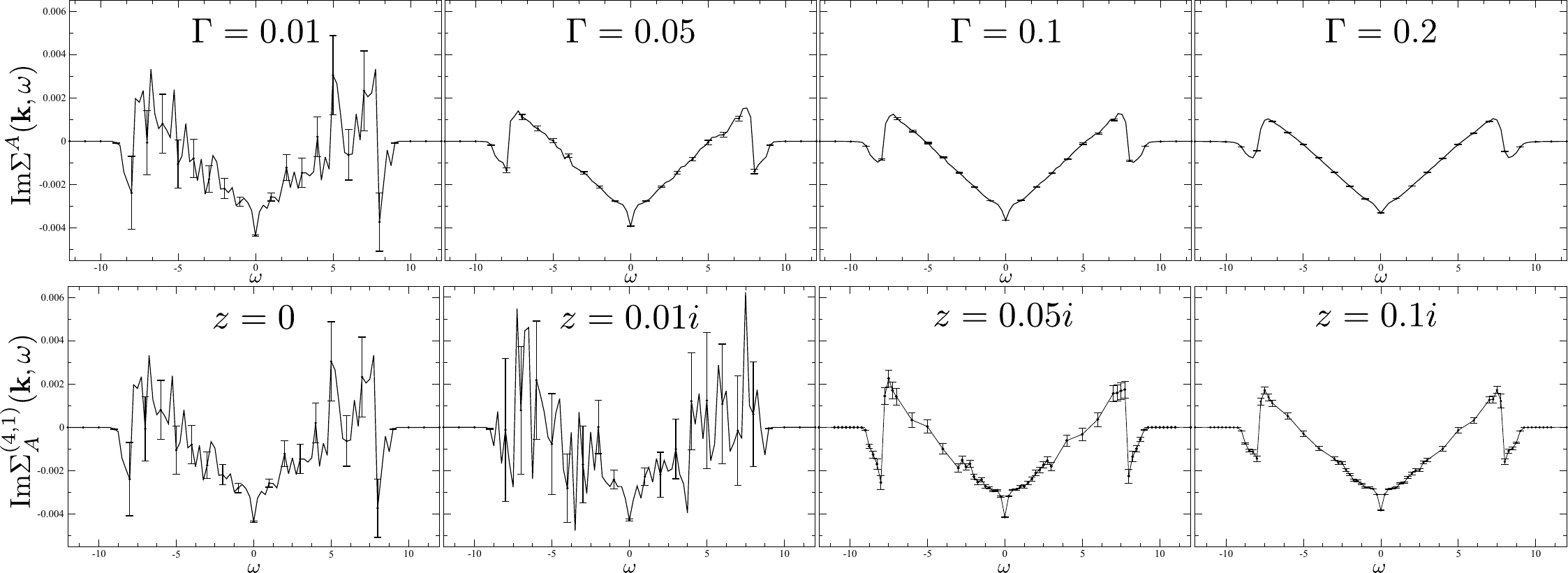}
    \caption{Evaluation of the imaginary part of the self energy evaluated on graph A as a function of $\omega$ for various parameters. The general parameters are $U/t=1$, $\beta t=5.00$, $\textbf{k}=(\pi, 0)$. Top row: Regular approach of having no counter-terms and $z=0$ (no effective chemical potential shift) for various $\Gamma$. Bottom row: Proposed approach of using renormalization shifts of varying purely imaginary values $z$ with $\Gamma/t=0.01$, summing all 8 diagrams up to 1 insertion.}
    \label{fig:matchbook}
\end{figure*}
In Fig.(\ref{fig:matchbook}) (top row) we evaluate the root diagram of graph $A$, $\textrm{Im}\Sigma^A(k,\omega+i\Gamma)$, with no insertions and $z=0$ for various values of $\Gamma$. We see that as $\Gamma$ is increased the error bars decrease but also the sharp features of the plot smooth out. This smoothing is not physical, but an artifact of a finite value of $\Gamma$.   One can see that while there are real frequencies that do not have a strong $\Gamma$ dependence, there are some $\omega$ where the result for $\textrm{Im}\Sigma$ varies significantly with $\Gamma$, most notably $\omega=0$ and $\omega=8t$ (near the band edge). For $\omega=0$, the $\textrm{Im}\Sigma^A(\textbf{k}, \Gamma)$ varies by $\approx11\%$, $15\%$, $26\%$ when $\Gamma$ increases to $\Gamma=0.05, 0.1, 0.2$ respectively in reference to the $\textrm{Im}\Sigma^A(\textbf{k}, \Gamma=0.01)$ value.

In the bottom row of Fig.(\ref{fig:matchbook}), the renormalized approach is used with the smallest broadening ($\Gamma=0.01$) from the above row, summing contributions from graph A up to 1 insertion. From these plots, one is able to see the importance of the choice of $z$ as too small of a shift leads to summing extra diagrams with a small overall broadening ($z=0.01i$), but for a large enough $z$, the broadening makes the extra diagrams tolerable giving smaller error bars and most importantly, preserving the features of the plot for all values of $\omega$ since a small $\Gamma$ is used. For comparison, at $\omega=0$, $\textrm{Im}\Sigma_A^{(4,1)}(\textbf{k}, z)$ varies by $\approx 4\%$, $5\%$, $11\%$ for $z=0.01i, 0.05i, 0.1i$ respectively in reference to the $\textrm{Im}\Sigma_A^{(4,1)}(\textbf{k}, z=0)$ value. 
Here the inclusion of only a single order of counter term diagrams produces an accurate result that preserves the sharp details of the $\Gamma\to 0^+$ limit. 

\subsection{Comparison of Computational Effort}\label{sec:comptime}
\begin{figure}
    \centering
    \includegraphics[width=\linewidth]{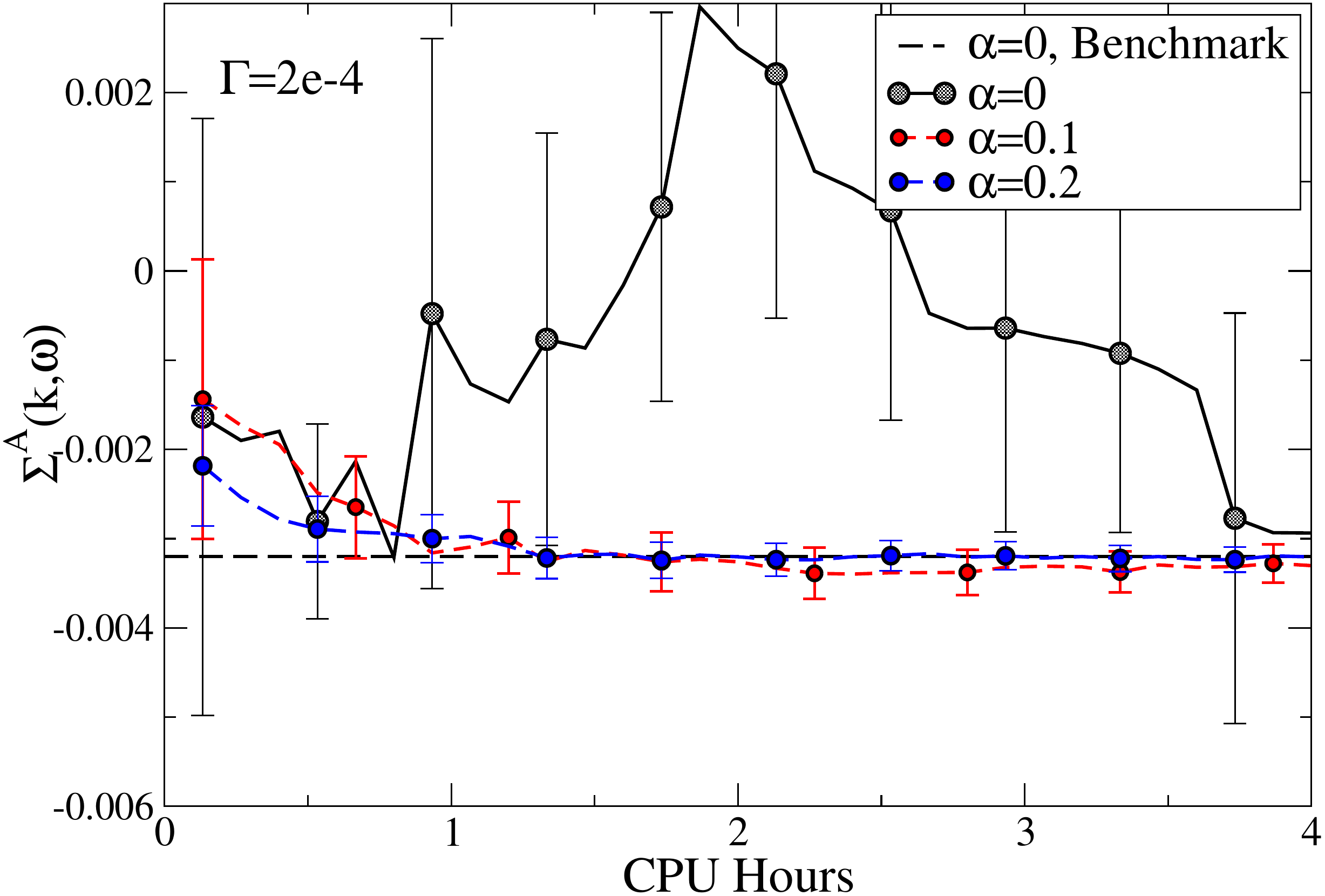}
    \caption{Evaluation of graph $A$ at $\beta t=5$, $k=(0,\pi)$, at $\omega/t=0.3$, $\Gamma/t=2\times 10^{-4}$.  Time for evaluation is for a single cpu.  In the case of $\alpha=0$ only the single root diagram is computed, while $\alpha \neq 0$ is evaluation of 36 diagrams up to $c=2$ which combined take the time as shown.}
    \label{fig:time}
\end{figure}

We have thus-far demonstrated that it is possible to obtain correct results within our renormalized approach.  However, the true power of this method becomes apparent when working with extremely small values of $\Gamma$ where the direct evaluation of the root diagram is virtually impossible.  We give such an example in Fig.~\ref{fig:time} where we contrast results for a difficult case with a value of $\Gamma=2\times 10^{-4}$ (50 times smaller than in the lower frame of Fig.~\ref{fig:matchbook}) for the cases of $\alpha=0,0.1,$ and 0.2.  In the case of $\alpha=0$ we compute only the root diagram while for $\alpha \neq 0$ we compute the root diagram plus 35 counter-term diagrams when truncated at $c=2$.  Plotted with uncertainties as a function of computational time, we see that for finite $\alpha$, despite having to sum more diagrams, the result converges quickly while on the same scale the $\alpha=0$ case is wildly inaccurate.  This inaccuracy is due to the dimensionality of the integrand (eight spatial dimensions) compounded by the sparse nature of the function in the small $\Gamma$ limit.  The $\alpha=0$ case does eventually converge which we show as the dashed-black curve which here represents $\approx 640$ cpu-hours.  This makes clear the need for our renormalized method if one wants to correctly approach the $\Gamma \to 0^+$ limit with reasonable computational effort. 

\section{Conclusions}
Real frequency DiagMC methods allow for true analytic continuation of Feynman diagrams which in principle alleviates the need for ill-posed methods of numerical analytic continuation.  These RF-DiagMC methods are not without their own difficulties in that they produce extraordinarily complicated analytic expressions comprised of many large in amplitude, but also largely cancelling, terms.  This issue is exacerbated by the $\Gamma \to 0^+$ limit in analytic continuation where numerical integration of remaining degrees of freedom will fail in general and this failure grows exponentially with perturbative order.  

We have demonstrated a scheme, equivalent to renormalized perturbation theory, whereby the exponential growth of complexity is squashed by including a complex renormalization that simultaneously acts as a regulator.  The price one pays for this exponential speedup is that one must compute a potentially large number of counter-term diagrams.  We have demonstrated that accurate results can be obtained so long as the regulator $z=i\alpha$ remains restricted to values less than the first Fermionic Matsubara frequency.  Towards zero temperature the utility of this approach is therefore expected to fail.  Nevertheless, the value of $\alpha$ can be tuned sufficiently small so as to require only a handful of low-order counter-term diagrams, in which case the computational advantage is potentially massive.  This is particularly the case in existing renormalization schemes where one \emph{already} computes the counter-term diagrams and this approach has no additional computational expense.\cite{leblanc:2022,chen:2019}

\acknowledgements{
JPFL acknowledge the support of the Natural Sciences and Engineering Research Council of Canada (NSERC) (RGPIN-2022-03882). 
Computational resources were provided by ACENET and the Digital Research Alliance of Canada. Our Monte Carlo codes make use of the open source ALPSCore framework   \cite{gaenko:2017} \cite{alpscore_v2}.}

\bibliographystyle{apsrev4-2}

\bibliography{refs.bib}

\end{document}